\documentclass[%
 reprint,
 amsmath,amssymb,
 aps,
prb,
]{revtex4-2}

\usepackage[version=3]{mhchem} 
\usepackage{tikz,xcolor}
\usepackage[hidelinks]{hyperref}
\usepackage{physics}
\usepackage{mathtools}
\usepackage[utf8]{inputenc}
\usepackage{color}
\usepackage{enumerate}
\usepackage{chemformula}[2014/04/08]
\usepackage{multirow}
\usepackage{tabularx}
\usepackage{indentfirst}
\begin{document}
\setchemformula{kroeger-vink}

\preprint{APS/123-QED}

\title{Molybdenum Defect Complexes in Bismuth Vanadate}

\author{Enesio Marinho Jr\href{https://orcid.org/0000-0003-4040-0618}{\includegraphics[scale=.05]{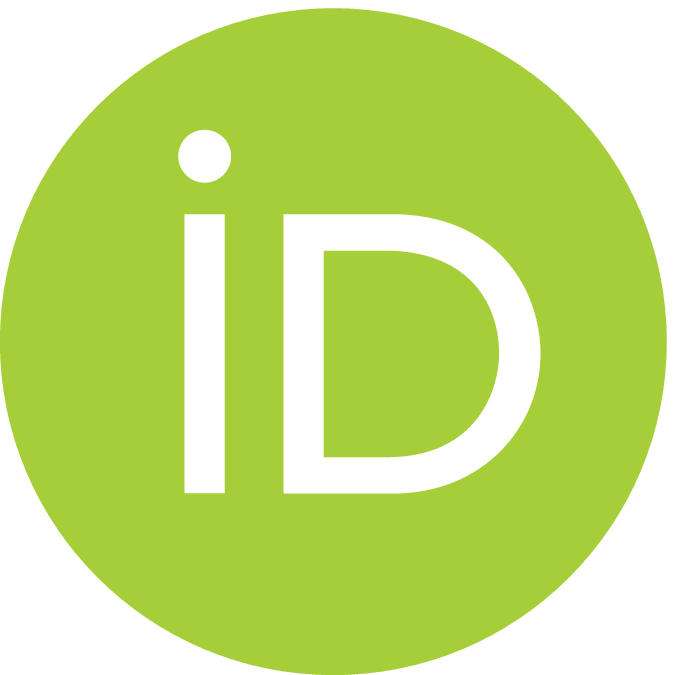}}}
\email{enesio.junior@ufabc.edu.br}
\author{Cedric Rocha Leão\href{https://orcid.org/0000-0002-1365-6514}{\includegraphics[scale=.05]{orcid_id}}}
\email{cedric.rocha@ufabc.edu.br}
\affiliation{Federal University of ABC (UFABC), 09210-580 Santo André, São Paulo, Brazil. 
}%




\date{\today}

\begin{abstract}
 Monoclinic bismuth vanadate (\ch{BiVO4}) is a promising $n$-type semiconductor for applications in sunlight-driven water splitting. Several studies have shown that its photocatalytic activity is greatly enhanced by high concentrations of Mo and W dopants. In the present work, we performed \textit{ab initio} calculations to assess the most favorable relative position between \ch{Mo}-related pairs in \ch{BiVO4}.
Surprisingly, we verify that the lowest energy configuration for \ch{Mo_V} pairwise defects in \ch{BiVO4} occurs on nearest-neighbor sites, despite the higher electrostatic repulsion and larger  strain on the crystal lattice. Similar results were observed for \ch{W_V} defect pairs in W-doped \ch{BiVO4}. We show that the origin of this effect lies in a favorable hybridization between the atomic orbitals of the impurities that is only verified when they are closest to each other, resulting in an enthalpy gain that overcomes the repulsive components of the pair formation energy. As a consequence, Mo and/or W doped \ch{BiVO4} are likely to present donor-donor defect complexes, which is an outcome that can be applied in experimental approaches for improving the photocatalytic activity of these metal oxides.  
\end{abstract}

\maketitle

\section{\label{sec:intro}Introduction}
Semiconductor metal oxides have been intensively investigated for photoanodes in photoelectrochemical (PEC) water splitting cells \cite{fujishima1972}. Promising materials for energy conversion and storage through water splitting must present chemical stability, relatively low cost, suitable  band  edge  positions, high optical absorption, long lived excitations and large carrier drift lengths \cite{Cooper2016, YatLi2017}. Metal oxides are usually resilient to water corrosion and are low cost materials, but also present crucial intrinsic limitations. For example, \ch{TiO2} with a wide band gap has low efficiency in absorbing visible light, \ch{Fe2O3} despite the moderate band-gap, has unfavorable band edge alignment relative to water's electrolysis potentials \cite{YatLi2017, AbdiReview2017}. 

Monoclinic scheelite-type bismuth vanadate (\ch{BiVO4}) has emerged as a promising complex metal oxide photoanode, since Kudo \textit{et al.} \cite{kudo1998} first reported its high visible light photoactivity. \ch{BiVO4} has been estimated theoretically to have a potential to harvest up to 11\% of the solar spectrum, delivering a photocurrent of 7.5 mA/cm$^2$, with  9\% solar-to-hydrogen conversion efficiency under AM 1.5 sunlight illumination \cite{abdi2013}. This owes to \ch{BiVO4} moderate band gap, $2.4$$-$$2.5$ eV \cite{kudo1999, Cooper2015} and good band edge's alignment with respect to  water's redox potentials. Its conduction band edge is close to 0 V $vs$ RHE \cite{YatLi2017, Walsh2009}, requiring low applied external bias to drive PEC's hydrogen production. 

Despite these promising properties, the experimental performance of pristine \ch{BiVO_{4}} as photoanodes is significantly limited by some key factors, such as short charge carrier diffusion lengths and high electron-hole recombination rate \cite{YatLi2017, AbdiReview2017, sivula2016}. 

Different experimental approaches have been adopted to overcome these limiting factors of \ch{BiVO4}, including crystal morphology control \cite{zhao2017}, heterojunctions \cite{ye2019, Zhiqiao2014}, tandem PEC devices \cite{abdi2013tandem}, coupling with oxygen evolving catalysts \cite{Kim990OEC, abdi2012nature}, and extrinsic doping \cite{parmar2012, wang2013, kim2015simultaneous}.

Doping is possibly the simplest attempt to improve the performance of metal oxides in PECs by supplying additional free carriers. \ch{BiVO4} has been effectively $n$-type doped with tungsten  (W)  and  molybdenum (Mo) \cite{Yin2011,Park2011}.  When \ch{Mo^{6+}}
or \ch{W^{6+}} is embedded into \ch{V^{5+}} sites, even at moderate concentrations,  the photoelectrochemical performances of the doped \ch{BiVO4} electrodes has been noticeably enhanced \cite{luo2013pccp}. Jeong \textit{et al.} \cite{jeong2013JPCC} have experimentally demonstrated that there exists an optimal doping concentration of \ch{Mo} and \ch{W} in \ch{BiVO4} that maximizes the performance of the photoanode.
The reported optimal concentrations are fairly high (8\% and 10\%, respectively), indicating that a comprehensive study of complexes involving these dopants in \ch{BiVO4} is important to rationalize and potentially enhance their beneficial effects for PEC hydrogen production.
%
Recently, Pakeetood \textit{et al.} \cite{pakeetood2019hybrid}  have reported that \ch{Mo_V} defects do show a tendency to group  with \ch{W_V} defects in co-doped material, giving rise to donor-donor complexes. The authors argue that this is an unexpected trend owing to the repulsive coulomb interaction between the donor defects. 

In the present work, we studied how Mo-defects interact with each other in a \ch{BiVO4} matrix. Using density functional theory we investigated how the electronic properties of \ch{BiVO4} with pairs of Mo substitutional dopants depend on the relative position between these impurities. Applying the well established formalism of first-principles calculations for point defects in solids, combined with analysis of the electronic properties, we also observe that pairs of Mo-defects present lowest formation energy on nearest-neighbor sites. The formation energy rises for intermediate distance between the dopants and then drops again as they get farther apart. We show that this surprising behavior can be rationalized by competing effects between local lattice strain, electrostatic repulsion, and gain of enthalpic stability through hybridization of the electronic clouds of the defects achieved at short distances. 


\section{\label{sec:method} Computational Details}

Our \textit{ab initio} calculations were based on the density functional theory \cite{HohenbergKohn1964, KohnSham1965}, as implemented in Vienna \textit{ab initio} simulation package (\textsc{vasp}) \cite{VASP}. The projector augmented wave (PAW) method \cite{PAW} were used to treat the electron-ion interaction,  and the exchange-correlation energy was described by the generalized-gradient approximation (GGA) as proposed by Perdew, Burke, and Ernzerhof (PBE) \cite{PBE}.

Structural optimizations were achieved using conjugate gradient algorithm until the Hellmann-Feynman forces on all atoms reach values lower than or equal to $0.025$ eV/\AA. Kohn-Sham orbitals were expanded into a  plane-wave  basis  set with a cutoff energy of $500$ eV. The Bi 5$d$\,6$s$\,6$p$, V 4$s$\,3$d$, O 2$s$\,2$p$, and Mo 4$s$\,4$p$\,4$d$\,5$s$ electrons were treated as valence electrons.

We consider the base-centered monoclinic primitive cell containing 2 units of \ch{BiVO_{4}} (12 atoms), with symmetry described by the standard space group C2/$c$. The optimized lattice parameters for the conventional unit cell were $a = 7.325$ \AA, $b= 11.765$ \AA, $c=5.179$ \AA, and $\beta = 135.09^{\circ}$, which are in good agreement with previous theoretical and experimental reports \cite{wang2013,cheng-acta}. To study the Mo-related point defects, we considered a 216-atom supercell which was built up by a $3\times3\times2$ projection of the \ch{BiVO4} primitive cell. The Brillouin zone was sampled using a $\Gamma$-centered $7\times7\times7$ $\mathbf{k}-$point grid for the structural optimization of the \ch{BiVO_{4}} primitive cell, following the scheme proposed by Monkhorst and Pack \cite{Monkhorst-Pack}, and a reduced $\Gamma$-centered $3\times3\times4$  grid was used in the calculations with the 216-atom supercells. 

Experimental and theoretical preliminary results have indicated that the \ch{Mo} substitution in V sites is the most energetically favorable defect configuration for this impurity in \ch{BiVO4} systems \cite{Rettie2013, Yin2011}.
For this reason, we modeled the Mo-related point defects only as substituting Mo in V sites (\ch{Mo_V}). 

\begin{figure}[t!]
\centering
    \includegraphics[width=0.48\textwidth]{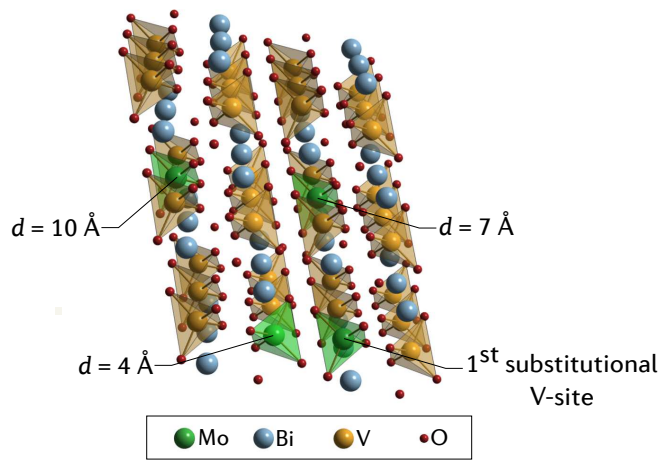}
\caption{Representation of the crystal structures of monoclinic \ch{BiVO4} doped with Mo. The green tetrahedrons highlight the \ch{V}-sites selected for the substitution with Mo atoms. For pair of \ch{Mo_V}, we considered three relative distances of the substitutional V-sites: 4 \AA, 7 \AA, and 10 \AA{} far apart from the first V-site. }
\label{fig:bvo-structure}
\end{figure}

First, we studied a single \ch{Mo_V} defect in \ch{BiVO4}. We analyzed the electronic structure and also the equilibrium formation thermodynamics of this point defect, besides the study of the intrinsic point defects considering the \ch{Bi}, \ch{V} and \ch{O} vacancies. Furthermore, we studied these same properties for pairs of \ch{Mo_V} defect into \ch{BiVO4}, examining the influence of the relative distances between substitutional V-sites in the electronic and structural properties, choosing substitutional sites $4$ \AA, $7$ \AA, and $10$ \AA~ far from the reference site (Fig.~\ref{fig:bvo-structure}).

\section{\label{sec:results-and-discussion}Results and Discussion}

\begin{figure*}[t!]
  \centering
    \includegraphics[width=\textwidth]{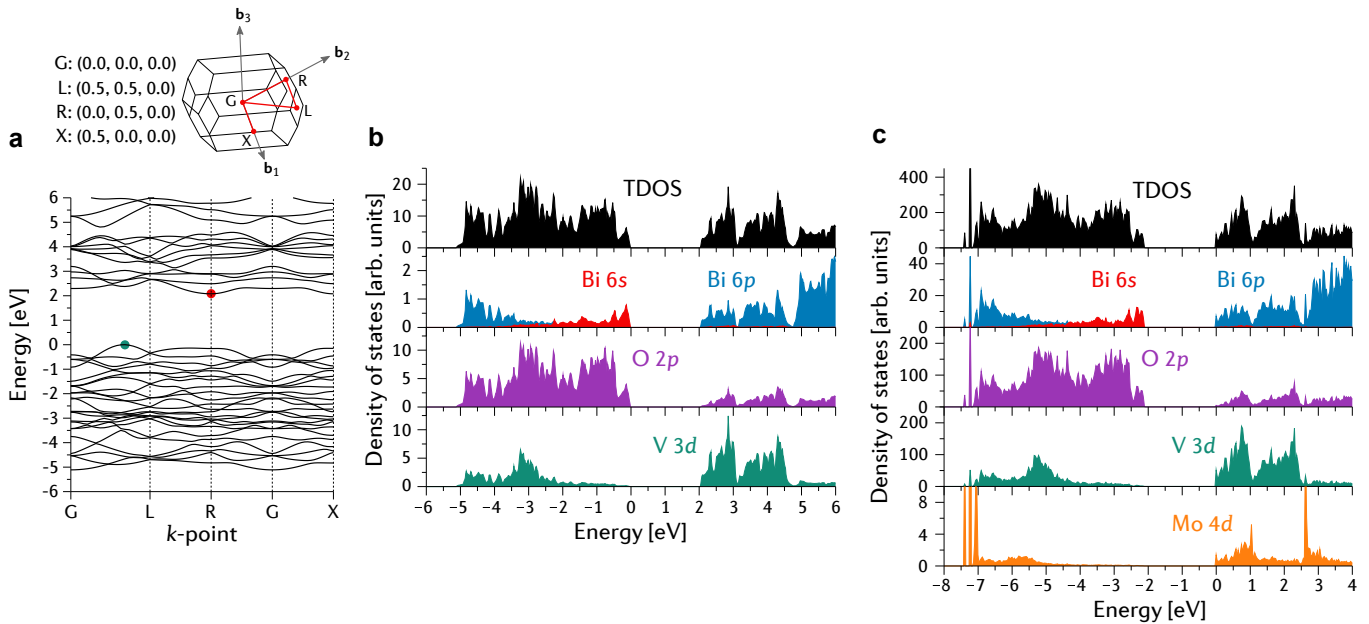}
    \caption{(a) Electronic band structure and (b) projected density of states of the \ch{BiVO4} primitive cell. Above the band structure, we depicted the selected high symmetry $k$-points in the Brillouin Zone. In (c), we presented the projected density of states of the 216-atom \ch{BiVO4} supercell with a single \ch{Mo_V} defect. The Fermi level was used as the zero-energy reference in each plot.}
    \label{fig:band-dos}
\end{figure*}

The GGA-PBE calculated band structure for the \ch{BiVO4} primitive cell is shown in Fig.~\ref{fig:band-dos}a. The result yields a bandgap of 2.08 eV, which is 0.40 eV underestimated relative to the experimental value \cite{cooper2014electronic}.  The band gap is indirect with CBM located at the $R$ point and the valence-band maximum (VBM) located in the $G$-$L$ direction. The direct gap is about 0.15 eV larger
than the indirect gap, in excellent agreement with previous experimental and theoretical reports \cite{Cooper2015, Walsh2009}.

Fig.~\ref{fig:band-dos}b shows the density of states (DOS) of the \ch{BiVO4} primitive cell, projected onto
the respective atomic orbitals. Both Bi $6s$ and O $2p$ orbitals contribute to the highest states of the valence band, whereas the V $3d$ orbitals form the states of the bottom of the conduction band. In Fig.~\ref{fig:band-dos}c, the orbital-resolved DOS of the 216-atom \ch{BiVO4} supercell containing a single \ch{Mo_V} is shown. The electronic contribution of orbital Mo $4d$ occurs mainly in deep energy levels of the valence band, 7 eV lower than the Fermi energy, and at the bottom of the conduction band. We do not notice significant changes in the shape of the valence and conduction band edges.

Defect formation can be seen as a thermodynamic process in which atoms and electrons are exchanged between the crystal containing point defects and the pristine one which acts as a particle reservoir. The analysis of point defect energies provides some fundamental information about the optoelectronic properties of semiconductors, such as if a certain defect tends to donate electrons to the system, accept electrons or remain neutral in the crystal. The occurrence of deep/shallow energy levels owing to these point defects can also be assessed.

The equilibrium defect concentration is calculated according to a Boltzmann distribution \cite{varotsos2013thermodynamics}:
{\small
\begin{equation}
\frac{n_{\text{eq}}}{N} 
\approx \exp\left(\frac{-\Delta H_{\text{f}}}{k_{\text{B}} T} \right),
\label{eq:concentration-defect}
\end{equation}
}\noindent where $n_{\text{eq}}$ denotes the number of point defects at equilibrium, $N$ is the total atomic sites involved in the defect formation, and $\Delta H_{\text{f}}$ is the enthalpy of formation. 

 
The formation enthalpy of a point defect in a charge state $q$ is given by
\begin{equation}
\Delta H_{\text{f}}= \left( E_{\text{D}} - E_{\text{p}} \right) + \sum_{i}n_i\,\mu_i + q(E_{\text{v}} + \mu_e ) + E_{\text{corr}}\,,
\label{eq:formEnergy}
\end{equation}

\noindent where $E_{\text{D}}$ and $E_{\text{p}}$ are the total energy of the supercell with the defect and of the pristine supercell, respectively; $\mu_i$ is the absolute value of the chemical potential of atom $i$; $n_i$ is the number of such defect atoms added ($n<0$) or removed ($n>0$); $q$ is the charge state of the defect; $\mu_e$ is the chemical potential of the reservoir with which the system exchanges electrons or holes (Fermi level); and $E_{\text{v}}$ is the valence band maximum energy.

Finally, the last term $E_{\text{corr}}$ corresponds to the finite size correction, which should be included to remove spurious electrostatic interactions between the charged defect and its periodic images \cite{PasquareloPhysRevB.86.045112, CedricPRB2011} . Makov and Payne \cite{makovPayne} (MP) described this correction energy, focusing on cubic cells, as follows:
\begin{equation}
E_{\text{corr}}^{\text{MP}} = \frac{q^2\alpha_{\text{M}}}{2\epsilon L}-\frac{2\pi qQ}{3\epsilon L^3}\,,  
\label{eq:makov-payne_Ecorr}
\end{equation}
\noindent where $L=\Omega^{-1/3}$ is the linear supercell dimension ($\Omega$ is the supercell volume), $q$ is the defect charge state, $\epsilon$ is the macroscopic dielectric constant of the medium, $\alpha_{\text{M}}$ is the appropriate Madelung constant for the respective supercell geometry, and $Q$ is the second radial moment of the localized charge distribution $\rho_{\text{c}}$:
\begin{equation}
    Q = \int_{V_{\textsc{sc}}} r^2\rho_{\text{c}}(\mathbf{r})\,d\mathbf{r}\,.
    \label{eq:Q_MP}
\end{equation}

We estimate the leading (first order) correction term in Eq. (\ref{eq:makov-payne_Ecorr}) for 216-atom monoclinic supercells by the Ewald method, computing the Ewald energy ($E_{\text{Ewald}}$) of a point charge (\ch{H^+}) placed into the supercell of interest and scaling the result by the calculated macroscopic dielectric constant. We obtained $E_{\text{Ewald}} = -1.454$ eV, and the calculated Madelung constant of \ch{BiVO4} supercell was $\alpha_{\text{M}} = 2.8$.  Furthermore, while $\epsilon$ strictly is a tensor, in our calculations we employed the lowest $\epsilon$ diagonal element. For \ch{BiVO4}, our results of
dielectric constants were $\epsilon_{xx} = \epsilon_{zz} = 7.24$ $\epsilon_0$ and 	$\epsilon_{yy} = 6.04$ $\epsilon_0$, in good agreement with other theoretical works \cite{wang2013, zhao2011}. Therefore, we adopted $\epsilon = 6.04$ $\epsilon_0$ in Eq. (\ref{eq:makov-payne_Ecorr}) aiming to apply the upper limit of this correction in the calculation of formation energies.

Lany and Zunger \cite{lany2008PRB, lany2009IOP} (LZ) have proposed to calculate  the  second radial moment in Eq. (\ref{eq:Q_MP}) considering that charge difference beyond the vicinity of the defect is predominantly described by a delocalized screening charge of density $n_{\text{s}}$ such that
\begin{equation}
   \rho_{\text{c}} \approx n_{\text{s}} =  \frac{q}{\Omega}\left(1-\frac{1}{\epsilon}\right) \,,
\label{eq:ns_LZ}
\end{equation}
and therefore the  second  radial  moment  could be calculated, substituting Eq. (\ref{eq:ns_LZ}) into Eq. (\ref{eq:Q_MP}). For a general geometry, with $a\neq b\neq c$ lattice parameters, we have that $Q = (1/12)(a^2 + b^2 + c^2)$. Using this result in Eq. (\ref{eq:Q_MP}), the image charge correction yields:
\begin{equation}
    E_{\text{corr}}^{\text{LZ}} = \left[1-c_{\text{sh}}\left(1-\frac{1}{\epsilon}\right)\right]\frac{q^2\alpha_{\text{M}}}{2\epsilon L}\equiv (1-f)\frac{q^2\alpha_{\text{M}}}{2\epsilon L}\,,
\end{equation}
in which the term $c_{\text{sh}}$ is the so-called shape factor. For instance, for a cubic cell $c_{\text{sh}}=\pi/3\alpha \approx 0.369$ \cite{PasquareloPhysRevB.86.045112, lany2009IOP}. For the 216-atom monoclinic supercell of \ch{BiVO4}, we obtain $c_{\text{sh}}= 0.419$, and this higher value was expected due to the anisotropic shape of the monoclinic supercell \cite{lany2009IOP}. 

Finally, we calculate the  image charge correction for 216-atom monoclinic supercell of \ch{BiVO4}, applying the LZ scheme: 
\begin{equation}
     E_{\text{corr}}^{\text{LZ}} \approx 0.651 \frac{q^2\alpha_{\text{M}}}{2\epsilon L}\,,
\end{equation}
and therefore we verified that although the monoclinic supercell is not approximately isotropic, the computed scaling factor $(1-f)$ is in excellent agreement with  $(1-f) \approx 2/3$ proposed by Lany and Zunger for systems approximately isotropic and with large macroscopic dielectric constants \cite{lany2008PRB, lany2009IOP}. 

To the thermodynamic study of the point defects, we considered as intrinsic defects the vacancies of Bi (\ch{V_{Bi}}), V (\ch{V_{V}}) and O (\ch{V_{O}}), and the extrinsic point defect considered was the Mo substitutional on V site (\ch{Mo_{V}}).

Defect formation energies are conventionally defined with respect to the chemical potential of the elemental solid, as shown by Eq.~(\ref{eq:formEnergy}). Numerical values of the atomic chemical potentials depend on the stoichiometric conditions under which the defects are created. In order to assure the stable growth of the desired compound, we must impose fundamental thermodynamic conditions to equilibrium chemical potentials, as detailed below \cite{PerssonPRB2005}: 
\begin{enumerate}[($i$.)]
\item To avoid precipitation of the atomic phases, the chemical potential of the atoms available to the crystal growth (the so-called \textit{atomic chemical potential}) should be smaller than the chemical potential of the respective elemental bulk or gas. That is:
\begin{equation}
\Delta \mu_{\text{Bi,V,O}} \equiv\mu_{\text{Bi,V,O}} - \mu_{\text{Bi,V,O}}^{\text{bulk/gas}}\leq 0\,.
\label{deltaMu}
\end{equation}
\item To maintain the thermodynamic stability of the \ch{BiVO4} crystal growth, the sum of the $\Delta\mu$ of the reacting elements must be equal to the heat of formation of \ch{BiVO4}: 
\begin{equation}
\Delta H_{\text{\ch{BiVO4}}} = \Delta \mu_{\text{\ch{Bi}}}+ \Delta \mu_{\text{\ch{V}}}+4\Delta \mu_{\text{\ch{O}}}\,, 
\end{equation}
\noindent where this heat of formation can be described as follows:
\begin{equation}
\Delta H_{\text{\ch{BiVO4}}} = \mu_{\text{\ch{BiVO4}}}^{\text{bulk}} - \left[\mu_{\text{\ch{Bi}}}^{\text{bulk}}+\mu_{\text{\ch{V}}}^{\text{bulk}}+4\mu_{\text{\ch{O}}}^{\text{gas}} \right]\,,
\end{equation} 
and each of these terms can be calculated by first-principles approach given the following definition of chemical potential:
\begin{equation}
    \mu^{\text{bulk/gas}} = \frac{E_{\text{total}}}{N_{\text{formulas}}}\,.
\end{equation}
\item The chemical potentials are further restricted by requiring that other possible competing phases are not formed. In the present work, we considered the following competing phases: Bi\textsubscript{2}O\textsubscript{3}, V\textsubscript{2}O\textsubscript{5} and VO\textsubscript{2}, as proposed in the Refs.~\cite{wang2013} and \cite{Yin2011}. Thus, we have that
\begin{eqnarray}
2\Delta\mu_{\text{Bi}}+3\Delta\mu_{\text{O}} &\leq& \Delta H_{\text{\ch{Bi_2O_3}}}\,,\\
\Delta\mu_{\text{V}}+2\Delta\mu_{\text{O}} &\leq& \Delta H_{\ch{VO_2}}\,,\\
2\Delta\mu_{\text{V}}+5\Delta\mu_{\text{O}} &\leq& \Delta H_{\text{\ch{V_2O_5}}}\,.\label{compPhase}
\end{eqnarray}

\begin{figure*}[t]
    \centering
    \includegraphics[width=\textwidth]{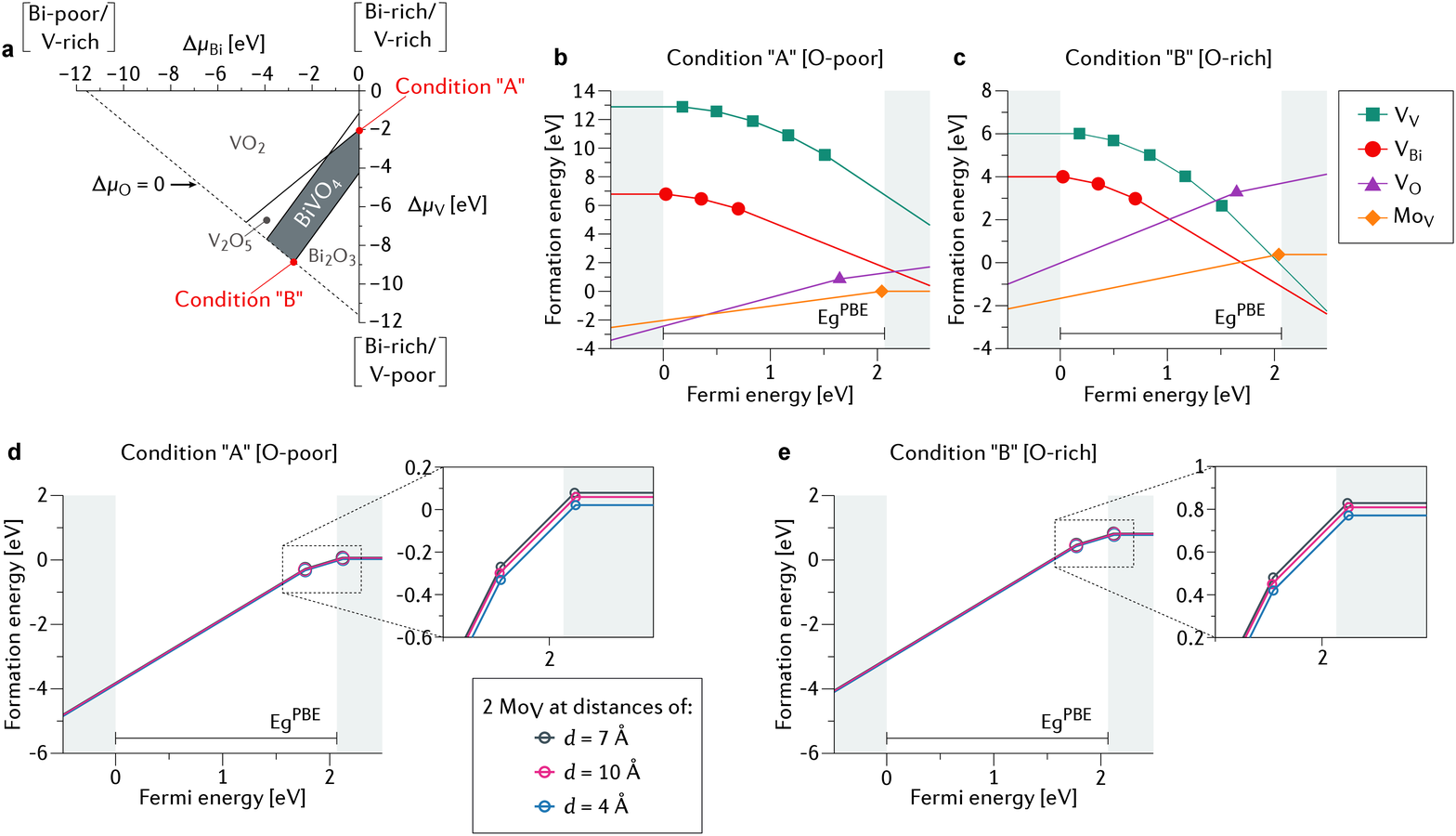}
    \caption{Accessible range of chemical potentials that stabilize the formation of \ch{BiVO4} and the formation energy of point defects under two growth condition. (a) The stable chemical potential region of \ch{BiVO4} in the ($\Delta\mu_{\text{Bi}}$, $\Delta\mu_{\text{V}}$) plane
with $\Delta\mu_{\text{O}} = 0$ eV (grey area). The conditions ``A''(\ch{O}-poor) and ``B'' (\ch{O}-rich) were chosen to numerically represent chemical environments for calculating the formation energy of the point defects. The curves in (b) and (c) show the formation energy of vacancies of vanadium (\ch{V_V}), bismuth (\ch{V_{Bi}}), oxygen (\ch{V_{O}}), and single substitutional \ch{Mo} in \ch{V}-site (\ch{Mo_{V}}) under conditions ``A'' (\ch{O}-poor) and ``B'' (\ch{O}-rich), respectively. In (d) and (e), the plots show the formation energy of pairwise substitutional \ch{Mo_V} defects considering three relative distances of the substitutional sites: 4 \AA, 7 \AA, and 10 \AA, under conditions ``A'' (\ch{O}-poor) and ``B'' (\ch{O}-rich), respectively.}
    \label{fig:formation-energy}
\end{figure*}

\begin{table*}[th!]
\caption{Transition energies $\epsilon(q/q')$ of \ch{Bi}, \ch{V}, and \ch{O} vacancies, as well as single and pairwise \ch{Mo_V} substitutional defects in \ch{BiVO4}. The energy values were calculated relative to the CBM for the donors, and to the VBM for the acceptors. 
\label{tab:Etransitions}}
\centering
\renewcommand{\arraystretch}{1.5}
\begin{tabular*}{0.7\textwidth}{l@{\extracolsep{\fill}}ccccc}
\hline 
Acceptor point defect & $\epsilon (0/ -1)$ & $\epsilon (-1/-2)$ & $\epsilon (-2/-3)$ & $\epsilon (-3/-4)$ & $\epsilon (-4/-5)$ \\ \hline
\ch{V_{Bi}} & 0.02 & 0.36 & 0.70 &   &   \\ 
\ch{V_{V}} & 0.18 & 0.50 & 0.84 & 1.17 & 1.51 \\  
Donor point defect & $\epsilon (0/ +1)$ & $\epsilon (+1/+2)$ & $\epsilon (+2/+3)$ & $\epsilon (+3/+4)$ & $\epsilon (+4/+5)$ \\
\ch{V_{O}} & 0.42&  &  &  &  \\ 
\ch{Mo_{V}} & 0.03 &  &  &  &  \\
2\ch{Mo_{V}} & $-0.05$  &  0.30 &  &  &  \\
\hline
\end{tabular*}
\end{table*}

\item Additional constraints must be posed to avoid the possible formation of compounds containing Mo and the forming elements of the host material. We have considered the following additional competing phases: MoO\textsubscript{2}, and MoO\textsubscript{3}:
\begin{eqnarray}
\Delta\mu_{\text{\ch{Mo}}}+2\Delta\mu_{\text{\ch{O}}} &\leq& \Delta H_{\text{\ch{MoO_2}}}\,,\\
\Delta\mu_{\text{\ch{Mo}}}+3\Delta\mu_{\text{\ch{O}}} &\leq& \Delta H_{\text{\ch{MoO_3}}}\,.\label{deltaMu_add}
\end{eqnarray} 
\end{enumerate}

Applying the thermodynamic constraints described in Eqs.~(\ref{deltaMu})-(\ref{deltaMu_add}), we achieved the accessible range for the atomic chemical potentials of \ch{Bi}, \ch{V}, and \ch{O}, which is depicted by the shaded area in the Fig.~\ref{fig:formation-energy}a. The formation energies of the vacancies and \ch{Mo}-related
point defects in \ch{BiVO4} were calculated using Eq.~(\ref{eq:formEnergy}), considering the conditions ``A'' (\ch{O}-poor condition) (Fig.~\ref{fig:formation-energy}b) and ``B'' (\ch{O}-rich condition).
Figs.~{\ref{fig:formation-energy}}b and {\ref{fig:formation-energy}}c show that the energies needed to form \ch{V_{Bi}} and \ch{V_{V}}
are higher than that to form \ch{V_O}, when $E_{\text{F}}$ is close to the top of the valence band (VBM). These differences tend to decrease considerably when $E_{\text{F}}$ is bordering the conduction band (CBM). This is in agreement with the fact that the highest valence band states are constituted mostly from O $2p$ orbitals, and the bottom of the conduction band is formed by V $3d$ states. In addition, \ch{V_{Bi}} and \ch{V_{V}} defects tend to be negatively charged throughout most of the allowed range Fermi level can assume. \ch{V_{O}} defects tend to be positively charged. For this reason,  \ch{V_{Bi}} and \ch{V_{V}} can be described as hole-producing acceptors, and \ch{V_{O}} are electron-producing donors to the crystal. 

The position of the electronic state introduced by the defects relative to the band edges of the host material can be estimated by the transition energy $\epsilon(q/q')$, which is defined as the Fermi energy at which the charge state of a given defect spontaneously transforms from $q$ to $q'$ \cite{PerssonPRB2005}. The transition energies of the analyzed point defects in \ch{BiVO4} are shown in Table~\ref{tab:Etransitions}. These results were obtained based on the formation energy curves represented in Figs.~{\ref{fig:formation-energy}}b and {\ref{fig:formation-energy}}c. 
\ch{V_{Bi}} and \ch{V_V} defects are both deep acceptors, which means that their ionization energies are significantly above the thermal energy $k_{\text{B}}T$. The point defects \ch{V_O} and \ch{Mo_V} are, in turn, shallow donors, which means that they will be easily ionized and produce free electrons in the conduction band of the host material. These results are in agreement with other theoretical reports \cite{Yin2011, wang2013} as well as experimental observations \cite{jeong2013JPCC}.

As described by Eq.~({\ref{eq:concentration-defect}}), the point defect concentration varies with the negative exponential of the formation energy. That is, the lower is the formation energy of a given point defect, the higher will be the prevalence of this defect in the crystal. Thus, considering the expected PEC properties of the \ch{Mo}-doped \ch{BiVO4} photoanodes, the optimal thermodynamic condition in which we have high concentration of shallow donor defects and low concentration of deep acceptors is depicted by the condition ``B'', with an \ch{O}-poor crystal growing environment. The highest concentration of donor defects in photoanodes introduces larger amounts of free electrons in the conduction band improving the photoelectrochemical performance.  On the other hand, deep levels, known as trap states are detrimental, since they capture photoexcited charge carriers, facilitating electron–hole recombination through Shockley-Read-Hall (trap-assisted) recombination \cite{park2018NRM}.

We also calculated the formation energy of pairs of \ch{Mo} substitutional defects. \ch{Mo_V}-\ch{Mo_V} defect pairs are shallow double donors with low formation energies. To investigate the interaction between these defects and the resulting variations in their electronic behavior we  
considered three relative distances  ($d_{\text{\ch{Mo}-\ch{Mo}}}$): 4 \AA, 7 \AA, and 10 \AA, under the growth conditions ``A'' (\ch{O}-poor) and ``B'' (\ch{O}-rich) (Figs.~\ref{fig:formation-energy}d and ~\ref{fig:formation-energy}e, respectively). 
Interestingly, the most stable pairwise defect configuration was achieved with the Mo dopants being incorporated into the nearest-neighbor substitutional V-sites, with $d_{\text{\ch{Mo}-\ch{Mo}}} = 4$ \AA{}. Seo \textit{et al.} \cite{seo2018} studied the interplay between the \ch{N} substitutional in \ch{O} sites (\ch{N_O}) and oxygen vacancies (\ch{V_O}) in \ch{BiVO4}, and their results have shown a energetic favorable tendency to form  \ch{N_O}-\ch{V_O} defect complexes in \ch{BiVO4}.  In most of the band gap region, \ch{V_O} defect exhibits $2+$ charge state, whereas \ch{N_O} defects present mostly $1-$ charge state. Hence, in this case the electrostatic attraction of the charged point defects can favor the formation of \ch{N_O}-\ch{V_O} defect complex.

For the \ch{Mo_V}-\ch{Mo_V} pair this result is surprising, since one would expect that the lattice strain induced by the defects would be largest when they are on nearest neighbor sites, leading to a higher formation energy. Similarly, the electrostatic repulsion between the positively charged donors should drive them apart.
We observe, however, that the formation energy for the defect complex is the lowest when they are on nearest-neighbor sites. For the intermediate distance configuration, it rises 58 meV (7.5\%) and then drops again 20 meV (-2.4\%) for the 10 \AA\ separation. This hints to the occurrence of some energy reduction effect associated to hybridization of the electronic clouds of the dopants in close proximity, balancing out the larger repulsion and greater strain of this configuration. As a consequence, this implies a tendency to form \ch{Mo_V} defect complexes.  

A similar behavior for Mo-W codeped \ch{BiVO4} has been reported recently by Pakeetood \textit{et al.} \cite{pakeetood2019hybrid}. The authors show that \ch{Mo_V}-\ch{W_V} defect complex is more likely to form than \ch{Mo_V}-\ch{V_{Bi}} or \ch{W_V}-\ch{V_{Bi}}, which are both donor-acceptor defects. To date, there is still a lack of physical understanding about these outcomes \cite{note:WVpairs}.


In order to understand this unexpected trend, we estimate the local stress field through the difference of the average distance between atoms in the defective and pristine supercell. We used one of the \ch{Mo} nucleus as the referential, for the defective supercells, and the corresponding \ch{V}-site as the referential for the pristine supercell. These distances were computed within a cutoff radius of 10.10 \AA, considering periodic boundary conditions. The sum of the distances, for the defective and the pristine supercell separately, was averaged by the total number of atoms within the cutoff radius as follows:
\begin{equation}
\langle d \rangle = \sum_{r_{\text{cutoff}}}d_{\text{atom}}/N_{\text{atoms}}\,,
\end{equation}
\noindent where $d_{\text{atom}}$ is the atomic spacing and $N_{\text{atoms}}$ is the total number of atoms in each evaluated supercell. Therefore, the local strain was estimated by the difference between $\langle d \rangle$ of the defective ($\langle d \rangle_{\text{2\ch{Mo_V}}}$) and pristine ($\langle d \rangle_{\text{pristine}}$) supercells: 
\begin{equation}
\langle\Delta d\rangle = \langle d \rangle_{\text{2\ch{Mo_V}}} - \langle d \rangle_{\text{pristine}}\,. 
\label{eq:avg-Deltad}
\end{equation}

Considering only the neutral charge \ch{Mo_V} defect pair, we obtained $\langle \Delta d\rangle$ equal to 0.072, 0.042, and 0.005 \AA/atom for $d_{\text{\ch{Mo}-\ch{Mo}}}$ of 4 \AA, 7 \AA, and 10 \AA, respectively. These results of $\langle \Delta d\rangle$ are exactly the same if we considered the charged systems, with $q=2+$. As expected, we notice that the increase in the separation between the substitutional sites results in a decrease in the local stress. Therefore, lattice strain indeed favors Mo defects farther apart, independently from other effects. As discussed above, we found that the lowest formation energy, however, happens when the two defects are incorporated on neighboring V-sites.

\begin{figure*}[t]
    \centering
    \includegraphics[width=\textwidth]{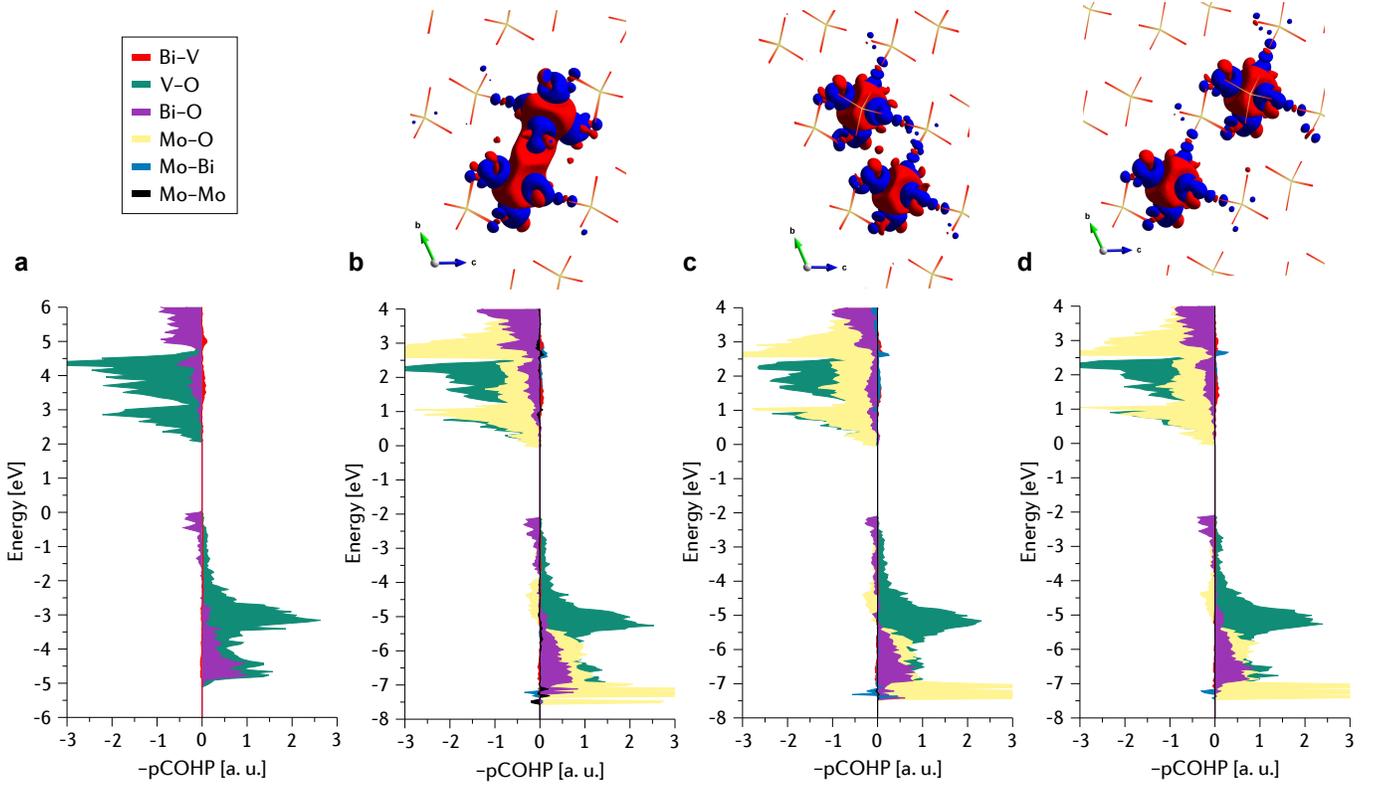}
    \caption{Crystal orbital hamiltonian populations of (a) \ch{BiVO4} pristine, or containing a double \ch{Mo_V} defect separated by (b) 4 \AA, (c) 7 \AA, and (d) 10 \AA~ from each other. The charge density differences for each \ch{Mo}-\ch{Mo} relative distances are shown above the respective COHP plots. Blue and red volumes represent electron accumulation and depletion, respectively, with isosurface of $0.005$ $e^-$\AA{}$^{-3}$. The Fermi level was set to zero in the energy axes. }
    \label{fig:cohp-chargedensity}
\end{figure*}

Our findings suggest that electronic effects are at play reducing the enthalpy of formation of the defect complex despite the largest strain and electrostatic repulsion at shortest distance.
To test this hypothesis, we first analyzed the charge density differences, which is shown in Fig.~\ref{fig:cohp-chargedensity} (upper panels). Since our purpose is to compare the interaction between the Mo atoms in the three configurations considered, we evaluated the charge distribution as follows:
\begin{equation}
\Delta\rho_{\text{\ch{BiVO4}}} = (\rho_{\ch{BiVO4}+2\ch{Mo_V}}) - (\rho_{\text{\ch{BiVO4}}}+ \rho_{\text{2\ch{Mo}}}+ \rho_{2\ch{V_V}})\,,
\end{equation}
\noindent in which $\rho_{\ch{BiVO4}+2\ch{Mo_V}}$, $\rho_{\ch{BiVO4}}$, $\rho_{2\ch{Mo}}$, and $\rho_{2\ch{V_V}}$ are the charge densities of the \ch{BiVO4} with a pair of \ch{Mo_V},  the pristine \ch{BiVO4}, the two Mo atoms isolated in the simulation box, and of the \ch{BiVO4} supercell with the \ch{V} vacancies (\ch{Mo} incorporation sites), respectively. The $\Delta \rho$ was calculated for the \ch{BiVO4} supercell with a pair of \ch{Mo_V} defects at distances of 4 \AA, 7 \AA, and 10 \AA~ from each other. The results were plotted using \textsc{vesta} code \cite{VESTA}.

The excess negative charge due to the \ch{Mo_V} defects acts as a perturbation of the local charge density distributions, being localized throughout the \ch{Mo}-\ch{O} tetrahedrons, with accumulation mainly around the \ch{Mo}-\ch{O} bond. We observe electron depletion close to the \ch{Mo} nuclei. Fig.~\ref{fig:cohp-chargedensity}b signalizes hybridization among the orbitals of the two Mo defects, with no indication of nodal points. This is not observed when the Mo defects are farther apart (Fig.~\ref{fig:cohp-chargedensity}c and Fig.~\ref{fig:cohp-chargedensity}d).

\begin{table*}[th!]
\caption{Integrated COHP values (in eV) of the respective filled interactions up to the Fermi level for the Mo-related interactions in \ch{BiVO4} with double \ch{Mo_V} defects in different substitutional-site distances $d_{\ch{Mo}-\ch{Mo}}$. 
\label{tab:iCOHP}}
\centering
\renewcommand{\arraystretch}{1.5}
\begin{tabular*}{0.7\textwidth}{l@{\extracolsep{\fill}}ccc}
    \hline
    \multirow{2}{*}{Interaction}&  \multicolumn{3}{c}{$d_{\ch{Mo}-\ch{Mo}}$}\\
     & 4 \AA&7 \AA&10 \AA\\
    \hline
\ch{Mo}-\ch{O}   & 2.864  & 2.734   & 2.792  \\
\ch{Mo}-\ch{Bi}   & -0.008    & -0.007   & -0.008 \\ 
\ch{Mo}-\ch{Mo}   & 0.005  & 0.000   & 0.000  \\ \hline
\end{tabular*}
\end{table*}

In order to establish on firmer grounds whether electronic hybridization is responsible for the observed tendency of \ch{Mo_V} to form stable pairs on neighboring sites, we performed a crystal orbital Hamiltonian population (COHP) analyses.  The energy-resolved visualization of chemical bonding in \ch{BiVO4} with a pair of \ch{Mo_V} was generated using the crystal orbital Hamiltonian population (COHP), as implemented in the \textsc{lobster} code \cite{LOBSTER, COHP1, COHP2, COHP3}. This involves a transformation of the plane wave basis set used by
\textsc{vasp}, to a localized basis set of Slater-type orbitals (STO) \cite{tao2019absolute}. The projected density of states is defined as:
\begin{equation}
    \text{PDOS}_i(E) = \sum_n |c_i^n|^2 \delta (E-E_n)\,,
\end{equation}
\noindent where $c_i^n$ are the coefficients of the molecular orbital expansion regarding the linear combination of atomic orbitals $\psi_n = \sum_i c_i^n \phi_i$. Thus, the COHP is
\begin{equation}
-\text{COHP}_{ij}(E) = H_{ij}   \sum_n c_i^n c_j^{*n} \delta (E-E_n)\,,
\label{eq:cohp}
\end{equation}
\noindent where $H_{ij}$ is the matrix element $\bra{\phi_i}H\ket{\phi_j}$ \cite{tao2019absolute}. The minus signal in the Eq. (\ref{eq:cohp}) is a mathematical artifact to describe bonding states as positive and anti-bonding states as negative results of the COHP projection. We used the following basis functions for the COHP calculations: Bi 5$d$6$s$6$p$, V 3$d$4$s$, O 2$s$2$p$, and Mo 4$s$4$p$4$d$5$s$, and analyzed the interactions between Bi-V, Bi-O, V-O, Mo-Bi, Mo-O, and Mo-Mo.

The COHP plot of pristine \ch{BiVO4} is shown in Fig.~\ref{fig:cohp-chargedensity}a. The curves reveal significant bonding interactions between \ch{V} and \ch{O} for the filled electronic states, showing that the scheelite structure of the \ch{BiVO4} is stabilized mostly by \ch{V}-\ch{O} bonds. On the other hand, we see bonding \ch{Bi}-\ch{O} interactions at energies around $4$ eV bellow the valence band maximum (VBM). We verify few antibonding states associated to \ch{Bi}-\ch{O} interactions near the VBM. Stoltzfus \textit{et al.} \cite{Stoltzfus2007} performed a study of the structure and bonding in different metal oxides, including \ch{BiVO4}, focusing only on \ch{Bi-O} interactions. Our COHP results agree with their reports for this interaction. Our analyses also show that the \ch{Bi}-\ch{V} interactions do not provide significant bonding/antibonding states in the valence band (Fig.~\ref{fig:cohp-chargedensity}a). 

In addition, we analyzed the integration of the Mo-related COHP up to the Fermi energy (ICOHP), which can provide useful quantification of Mo-related bond strength \cite{mann2019electrocatalytic, khazaei2019novel}. This furnishes a chemically intuitive interpretation of the electronic structure by comparing bond strength among the different elements in the structure \cite{PhysRevMaterials.3.094407}. The results of ICOHP are shown in the Table~\ref{tab:iCOHP}, in which the overall bonding and antibonding interactions up to Fermi level are represented by
negative and positive values, respectively. We observe a net bonding interaction of the Mo defects pairwise when they are in nearest-neighbor site. The bond strength of the Mo-Mo interaction vanishes when the defect pairs become farther apart. We also observe net anti-bonding states between Mo and Bi atoms which remain on similar levels in all cases. Therefore only in the nearest-neighbor configuration the net anti-bonding interactions are balanced out by Mo pairwise stabilizing interaction.      




\section{\label{sec:conclusion}Summary and Conclusions}
In summary, we investigated by first-principles calculations the electronic nature and the interaction between substitutional Mo defects in Mo-doped \ch{BiVO4}.  
Our analyses indicated that these shallow donors, that have been shown to enhance the photoelectrochemical performance of the material, display an unexpected inflection in their formation energy versus their relative distance curve. The most favorable configuration for \ch{Mo_V} defect pairs in \ch{BiVO4} is when they are on nearest-neighbor sites, despite the electrostatic repulsion they exert on each other. Our simulations also confirm that the lattice is more strained when the defects are on nearest-neighbor sites. This is also at odds with the observed lowest formation energy for this configuration. The COHP analyses showed that only in the nearest-neighbor configuration bonding states exist between the two Mo atoms, balancing out the electrostatic repulsion and lattice strain that would favor configurations with the defects farther apart. Our findings indicate that \ch{BiVO4} doped with Mo tends to form defect clusters, pointing out to new research issues that can improve the photoactivity of these metal oxides.






\section{\label{sec:acknowledgments}Acknowledgments}
This work was supported by the Brazilian Federal Agency for Support and Evaluation of Graduate Education (CAPES). We thank Lídia Carvalho Gomes and Juan Camilo Alvarez Quiceno for fruitful discussions. Computational resources
were provided by the high performance computing center at
UFABC. 


\begin{thebibliography}{54}
\bibitem [{not()}]{note:WVpairs}%
  \BibitemOpen
  \href@noop {} {}\bibinfo {note} {For completeness, we also calculate the
  formation energy of \ce{W_V} defect pairs, considering the same relative
  distances of the substitutional sites: 4 \AA{}, 7 \AA{}, and 10 \AA{}. The
  \ce{W} 5$p$5$d$6$s$ electrons were treated as valence electrons. We observe
  identical behavior between the defect pairs, with lowest formation energy for
  \ce{W_V} embedded in nearest-neighbor V-sites. Regarding the neutral charged
  \ce{BiVO4} with pairs of \ce{W_V}, for O-poor growth condition, we obtain
  $\Delta H_{\text{f}} =$ 0.68, 0.74, and 0.72 eV for relative distances of 4
  \AA{}, 7 \AA{}, and 10 \AA{}, respectively. Similarly, for O-rich growth
  condition, we obtain $\Delta H_{\text{f}} =$ 1.39, 1.45, and 1.43 eV for 4
  \AA{}, 7 \AA{}, and 10 \AA{}, respectively.}\BibitemShut {Stop}%
\end{thebibliography}

%
\end{document}